\documentstyle[preprint,prd,aps,floats,epsfig]{revtex}

\def\sech{{\rm sech}}
\def\tanh{{\rm tanh}}
\newcommand{\be}{\begin{equation}}
\newcommand{\ee}{\end{equation}}
\newcommand{\bea}{\begin{eqnarray}}
\newcommand{\eea}{\end{eqnarray}}
\newcommand{\bml}{\begin{mathletters}}
\newcommand{\eml}{\end{mathletters}}

\newcommand{\dal}                    % D'Alembertian
    {{\sqcap\!\!\!\!\sqcup\,}}

\begin{document}
\preprint{gr-qc/9903059}

\tighten

\renewcommand{\topfraction}{0.8}

\title{Thick self-gravitating plane-symmetric domain walls}
\author{Filipe Bonjour,\footnote{E-mail address:
          \texttt{Filipe.Bonjour@durham.ac.uk}} Christos
        Charmousis,\footnote{E-mail address:
          \texttt{Christos.Charmousis@durham.ac.uk}}
        and Ruth Gregory\footnote{E-mail address:
          \texttt{R.A.W.Gregory@durham.ac.uk}}}
\address{Centre for Particle Theory, 
         Durham University, South Road, Durham, DH1 3LE, U.K.}
\date{\today}
\maketitle

\begin{abstract}
  We investigate a self-gravitating thick domain wall for a $\lambda \Phi^4$
  potential. The system of scalar and Einstein equations admits two types of
  non-trivial solutions: domain wall solutions and false vacuum-de Sitter
  solutions. The existence and stability of these solutions depends on the
  strength of the gravitational interaction of the scalar field, which is
  characterized by the number $\epsilon$. For $\epsilon \ll 1$, we find a
  domain wall solution by expanding the equations of motion around the flat
  spacetime kink. For ``large'' $\epsilon$, we show analytically and
  numerically that only the de Sitter solution exists, and that there is a
  phase transition at some $\epsilon_{\rm max}$ which separates the two kinds
  of solution. Finally, we comment on the existence of this phase transition
  and its relation to the topology of the domain wall spacetime.
\end{abstract}

\section*{Introduction}

\noindent The spacetime of cosmological domain walls has now been a subject of
interest for more than a decade since the work of Vilenkin and Ipser and
Sikivie~\cite{Vil,IS}, who used Israel's thin wall formalism~\cite{Israel} to
compute the gravitational field of an infinitesimally thin planar domain wall.
This revealed the gravitating domain wall as a rather interesting object:
although the scalar field adopts a static solitonic form, the spacetime cannot
be static if one imposes a reflection symmetry around the defect~\cite{Vil,IS},
but displays a de Sitter expansion in any plane parallel to the wall. Moreover,
there is a cosmological event horizon at finite proper distance from the wall;
this horizon provides a length scale to the coupled system of the Einstein and
scalar field equations for a thin wall.

 After the original work by Vilenkin, Ipser and
Sikivie~\cite{Vil,IS} for thin walls, attempts focussed on trying to find a perturbative expansion in the wall thickness~\cite{RD,larryw}.
With the proposition by Hill, Schramm and Fry~\cite{HSF} of a late phase
transition with thick domain walls, there was some effort in finding exact
thick solutions~\cite{Goetz,Mukher}; however, these walls were supposed to be
thick by virtue of the low temperature of the phase transition, which means
that the scalar couples very weakly to gravitation. The suggestion that the
cores of defects created near the Planck time could undergo an inflationary
expansion~\cite{Vil2,Linde} then reopened the question of thick domain walls
(where now thick means relative to the wall's natural de Sitter horizon). This
time, the high temperature of the phase transition ensures that the scalar
field in this case interacts very strongly with gravity. Here, we consider
gravitating thick domain wall solutions with planar and reflection symmetry in
the Goldstone model; a more detailed discussion can be found in our
paper~\cite{BCG}, where we consider general potentials and de Sitter/anti-de
Sitter background spacetimes.

\section*{Domain walls in flat spacetime}

\begin{figure}[htbp]
  \begin{center}
    \epsfig{file=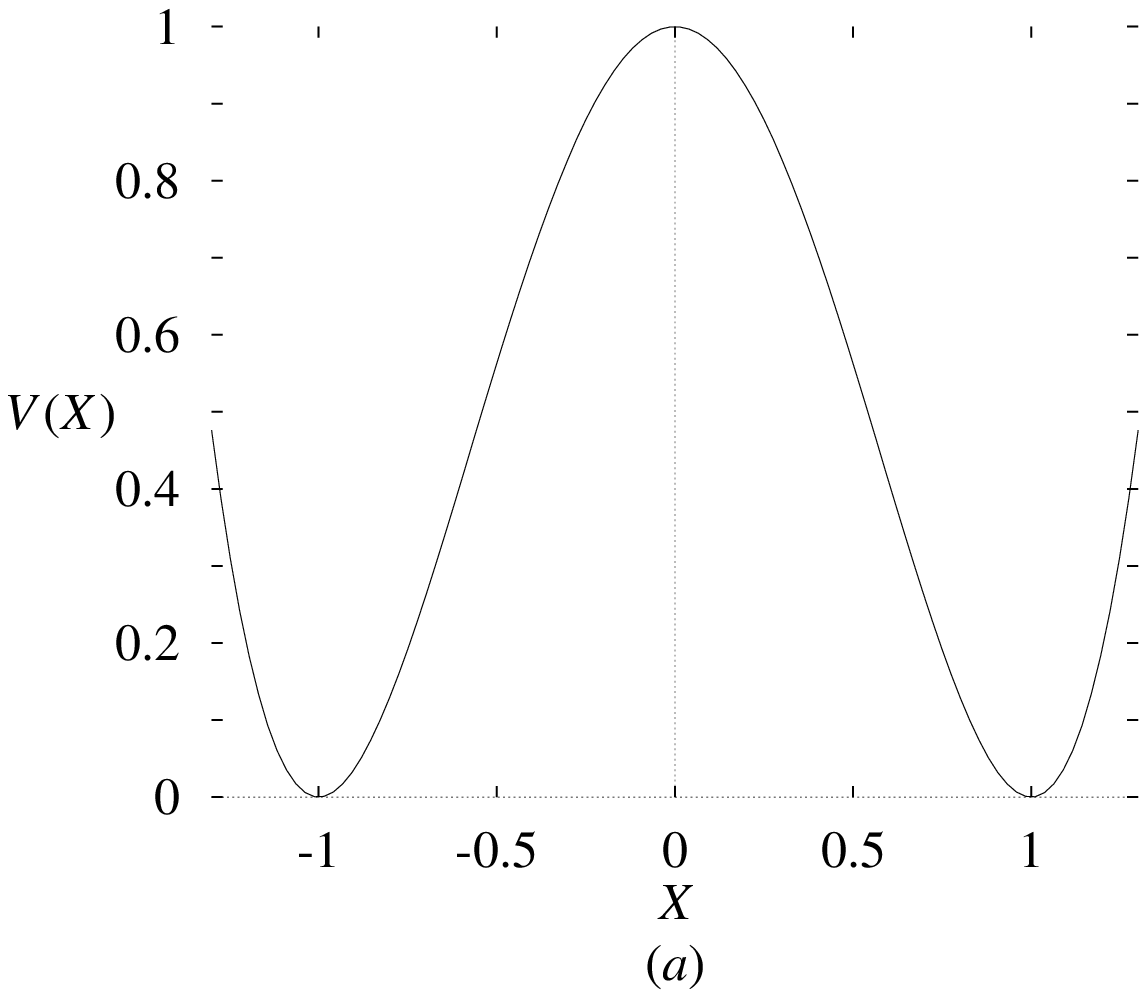,width=7cm}  \hfill
    \epsfig{file=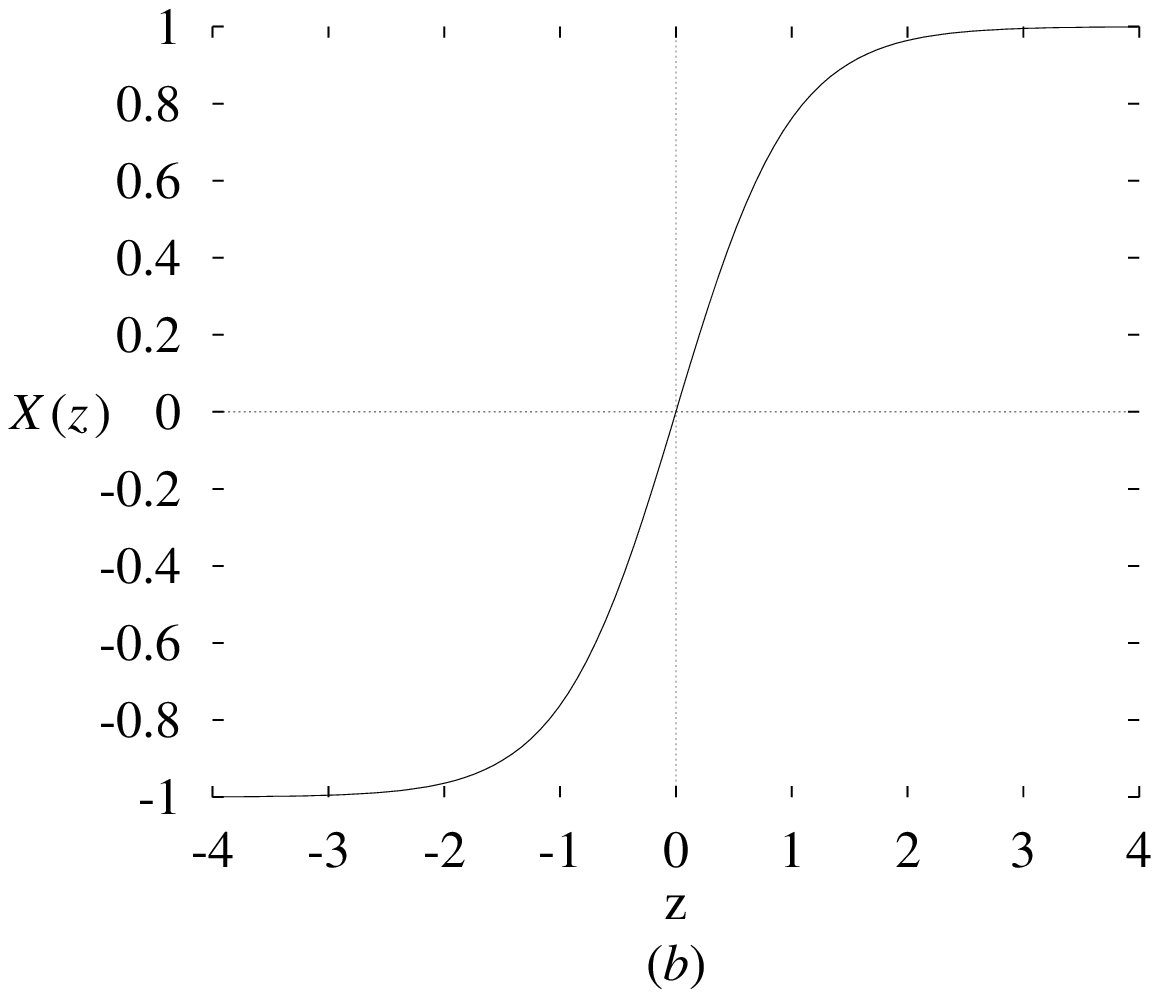,width=7cm}
  \end{center}
  \caption{({\it a\/}) The potential $V(X) = (X^2-1)^2$. ({\it b\/}) The flat
           spacetime kink solution, $X(z) = \tanh(z)$.}
  \label{fig:pot}
\end{figure}

In order to fix the notation let us first briefly review a domain wall in flat
space-time \cite{VS}. Consider a flat metric $\eta_{a b}$ with signature
$(+,-,-,-)$ and spacetime coordinates $x^a= \{t,x,y,z\}$. The matter Lagrangian
will be given by the $\lambda \Phi^4$ Lagrangian
\bea
  {\cal L} &=& \eta^{a b} \, \nabla_a \Phi \nabla_b \Phi - V(\Phi)
               \label{lag}\\
   V(\Phi) &=& \lambda \left ( \Phi^2 - \eta^2\right )^2, \label{pot}
\eea
where $\Phi=\Phi(z)$ is a real scalar field. For a potential with a non-trivial
degenerate set of minima, such as~(\ref{pot}), one gets domain wall solutions
because the vacuum manifold ${\cal M}=\{\pm \eta\}$ is not connected. We scale
out the dimensionful symmetry breaking scale $\eta$ by letting $X = \Phi/\eta$
and we set the wall's width $w ={1 \over \sqrt{\lambda} \eta}$ to unity (thus
measuring distances in wall units rather than Planck units). Then the matter
langrangian takes the simplified form
\be
  {\cal L} = -(X')^2 - (X^2-1)^2
\ee
with equation of motion
\be
 X''-2X(X^2-1)=0.
\ee
The well known kink solution is given by $X(z)=\tanh z$, an odd function
approaching exponentially the true vacua $\pm 1$ as $z \rightarrow \pm \infty$
(see figure~\ref{fig:pot}).

\section*{Gravitation and general setting of the problem}

Now let us couple our flat space-time domain wall solution to gravity in a
minimal way. Consider a domain wall with local planar symmetry, reflection
symmetry around the wall's core at $z=0$. The matter langrangian is given by,
\be
  {\cal L} =  g^{a b} \nabla_a X \nabla_b X - (X^2-1)^2 \\
\ee
and coupled to gravity via a metric admitting these symmetries, of components
$g_{a b}$. We suppose again that $X=X(z)$, a static field and the metric $g_{a
b}$ is given by,
\be
  ds^2 = A^2(z) \, dt^2 - B^2(z,t) \left( dx^2+dy^2 \right) - dz^2
\ee
with $z$ measuring the proper distance from the wall.

Now in order to find a domain wall solution we have to solve the coupled system
of differential equations consisting of the Einstein and scalar field equations
namely,
\bml\label{sys1} \bea
  R_{ab} &=& \epsilon \left[ 2X_{,a} X_{,b} - g_{ab} (X^2-1)^2 \right] \\
    \dal X + 2X(X^2-1) &=& 0,
\eea \eml
where $\epsilon=8\pi G\eta^2$ is a dimensionless parameter which we call
gravitational strength parameter and which characterises the gravitational
interaction of the Higgs field. Note that the Ricci tensor $R_{ab}$ is
generated by wall matter via the Einstein equations, which is the essence of
self-gravity.

Since the field profile function $X$ is time independent, one sees from the
Einstein equations that the metric reduces to,
\be
  ds^2 = A^2(z) \, dt^2 - A^2(z) e^{2kt} \left( dx^2+dy^2 \right) - dz^2.
\ee
Then~(\ref{sys1}) reduces to,
\bml \label{sys2} \bea
    {A''\over A} &=& -{\epsilon\over3} \left [ 2 X^{\prime 2} +
      (X^2-1)^2\right ] \label{sys2a} \\[2mm]
    X'' + 3 {A'\over A} X' &=& 2X(X^2-1) \label{sys2b} \\[2mm]
    \left ( {A'\over A} \right )^2 &=& {k^2\over A^2} + \epsilon
      \left [ X^{\prime 2} - (X^2-1)^2\right ] \label{sys2c}
\eea \eml
where the third equation is a constraint equation giving the value of the
constant $k$. One can see that~(\ref{sys2}) admits for all $\epsilon$ the false
vacuum de Sitter solution, $X = 0$, $A(z) = \cos kz$, $k^2 = \epsilon / 3$;
here, the role of the cosmological constant is played essentially by the 
parameter
$\epsilon$. For a domain wall solution we demand the following boundary
conditions,
\begin{itemize}
  \setlength{\itemsep}{0pt}
  \item $X(z)$ is an odd function (for a topological solution);
  \item $|X| \longrightarrow |X_0|<1, \quad \mbox{as} 
        \quad z\longrightarrow \pm z_{h}$;
  \item $X'(z_{\rm h}) = 0$;
  \item $A(0)=1, \qquad A'(0)=0$.
\end{itemize} 
We are expecting that in the presence of gravity our coordinate system will
break down at some $z_{\rm h}$, representing the proper distance of the wall's
core to the event horizon. This means that the scalar field will not fall all
the way down the potential to its minimum value, $\pm 1$, within the range of
validity of our coordinates. Moreover in order for the solution to be
non-singular we will have to suppose that $X'(z_{\rm h})=0$. The conditions on
$A(z)$ result from reflection symmetry and are always valid from the regularity
of the metric.

\section*{Perturbative solution for weak gravity}

When $\epsilon \ll 1$, corresponding to weak self-gravity, we can expand the
unknown fields in powers of $\epsilon$,
\bea
  X &=& X_0 + \epsilon \, X_1 + O \left( \epsilon^2 \right) \\
  A &=& A_0 + \epsilon \, A_1 + O \left( \epsilon^2 \right),
\eea
where to zeroth order we have of course the flat space-time solution,
\be
  X_0=\tanh(z), \qquad A_0=1.
\ee

\begin{figure}[htbp]
  \centerline{\epsfig{file=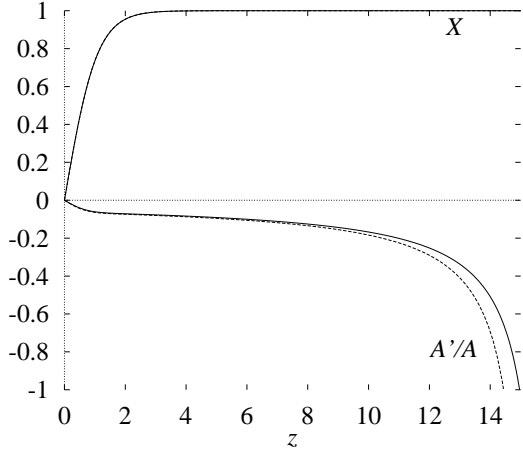,width=7.5cm}}
  \caption{Comparison between the numerical solution (solid lines) and the
           series solution~(\protect\ref{solser}). At this scale the two
           solutions for $X$ are superposed.}
  \label{fig:comp}
\end{figure}

The field equations~(\ref{sys2}) to first order in $\epsilon$ give
\bea
  A_1'' &=& -{1\over3} \left [ 2 X_0^{\prime2} + (X_0^2-1)^2 \right ]
             \label{onea}\\
  X_1'' &=& - 3 A_1' X_0' + 2X_1 (3X_0^2-1)\\
  k^2   &=& \phantom{-} \epsilon^2 \left [ A_1^{\prime2} + {2\over3} (X_1X_0''
            - X_0'X_1') \right ]
\eea
Solving in turn these equations we obtain,
\bml \label{solser} \bea
  X(z) &=& \tanh z - \frac \epsilon 2 \sech^2 z \left[ z + \frac13 \tanh z
        \right] + {\rm O} \left( \epsilon^2 \right) \\
  A(z) &=& 1 - \frac \epsilon 3 \left[ \ln \cosh z - \frac 12 \tanh^2 z
           \right] + {\rm O} \left( \epsilon^2 \right) \\
  k    &=& \pm{2 \over 3}\epsilon + {\rm O} \left( \epsilon^2 \right).
\eea \eml
Then since $A_1^{''}$ is negative, $A$ asymptotes $1-kz$ for large $z$. Thus at
proper distance from the wall given by, $z_{\rm h} = \frac1k \sim \epsilon^
{-1}$ we have an event horizon since $ A(z) \; \lower 5pt \vbox{ \hbox{
$\stackrel{ \longrightarrow}{\scriptscriptstyle z\to 1/k}$}} \; 0$ and
$g_{tt}=A^2$. Figure~\ref{fig:comp} compares the perturbative
solution~(\ref{solser}) with the numerical solution.

\section*{Strongly self-gravitating walls}

Let us now consider the case of strong self-gravity, corresponding to $\epsilon
= {\rm O}(1)$, for supermassive walls forming near the Planck scale. For such
large $\epsilon$ where $\eta \sim M_{\rm Pl}$ our perturbative analysis breaks
down. Nevertheless generically one can consider that since for smaller values
of $\epsilon$, $z_{\rm h} \sim \epsilon^{-1}$,

\vspace{\baselineskip} \centerline{bigger $\epsilon \qquad \longleftrightarrow
\qquad$ smaller $z_{\rm h}$.} \vspace{\baselineskip}

\begin{figure}[htbp]
  \begin{center}
    \epsfig{file=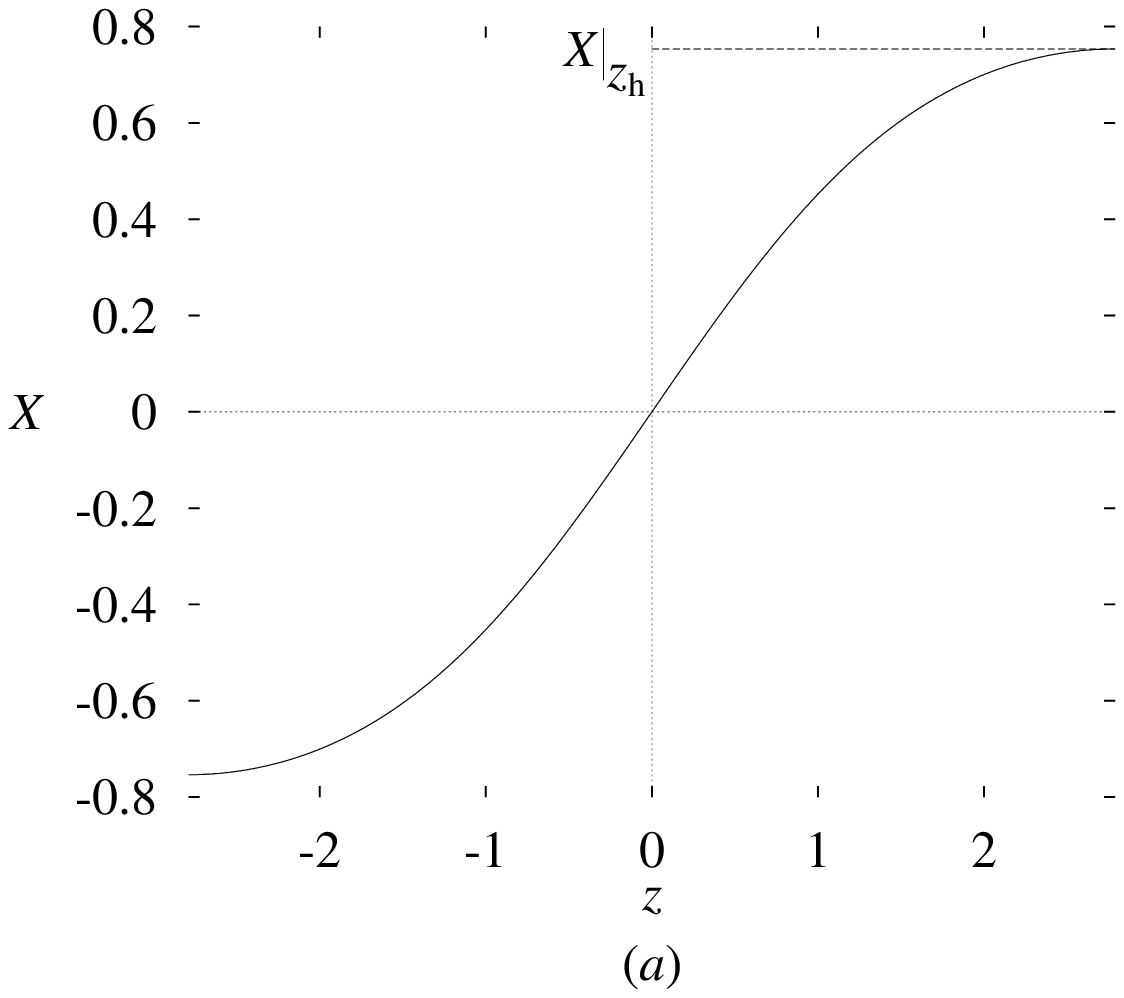,width=7cm} \hfill
    \epsfig{file=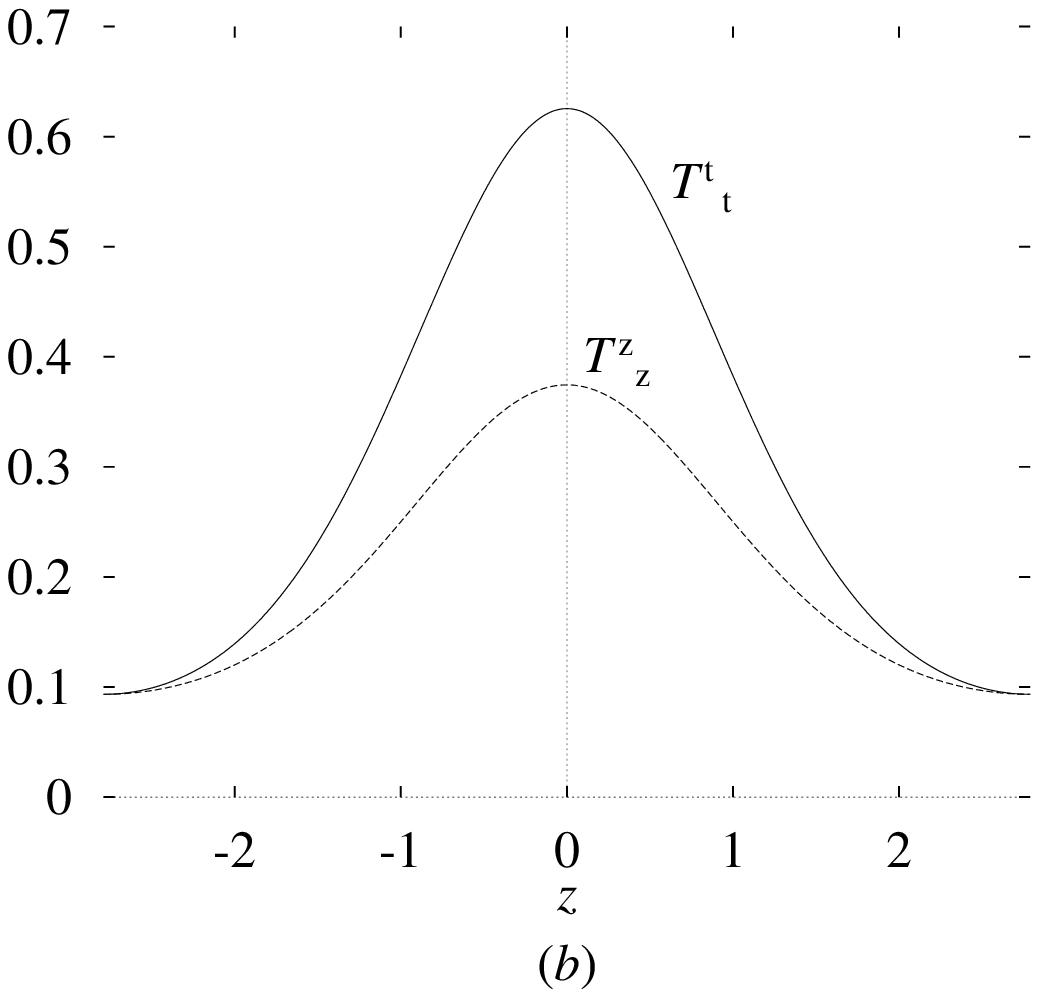,width=7cm} \\[3mm]
    \epsfig{file=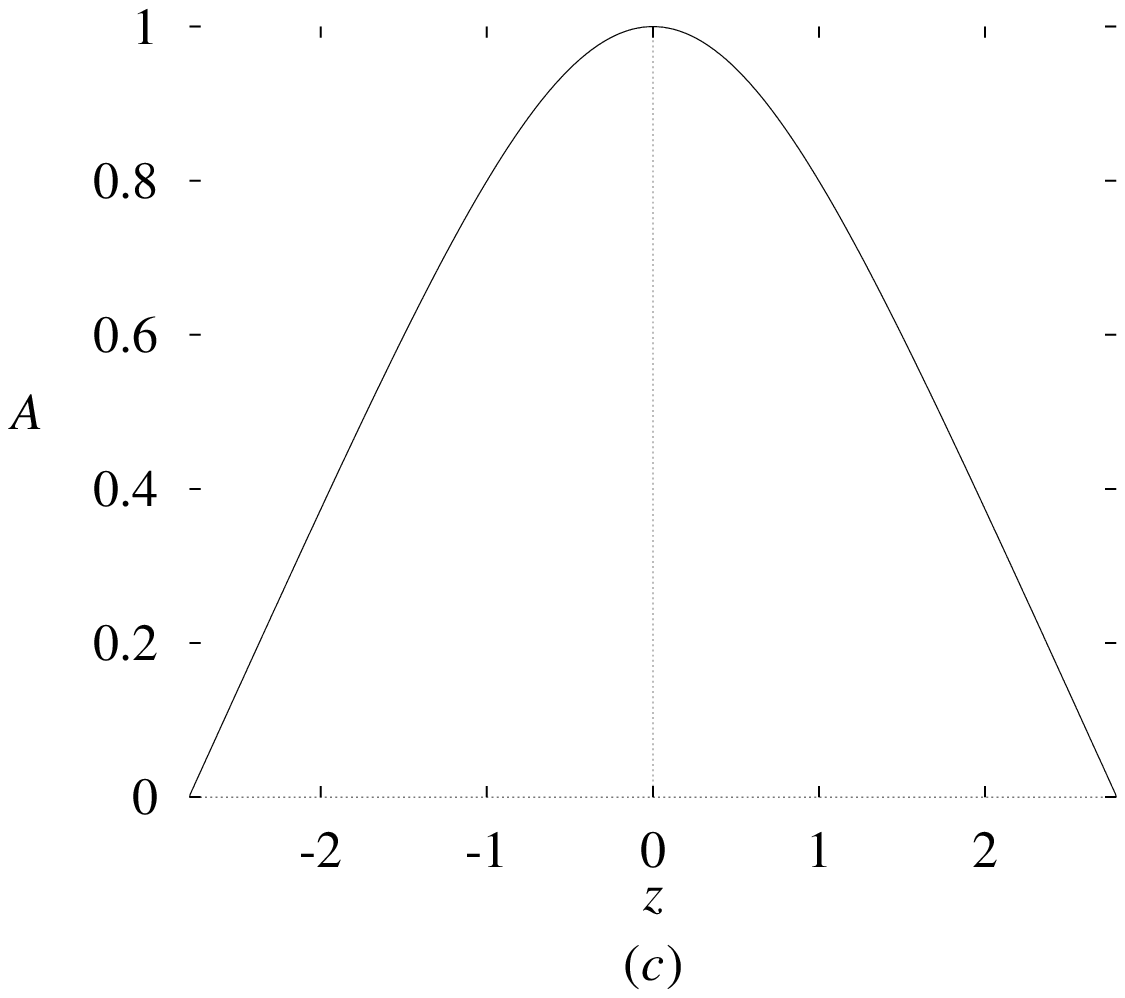,width=7cm} \hfill
    \epsfig{file=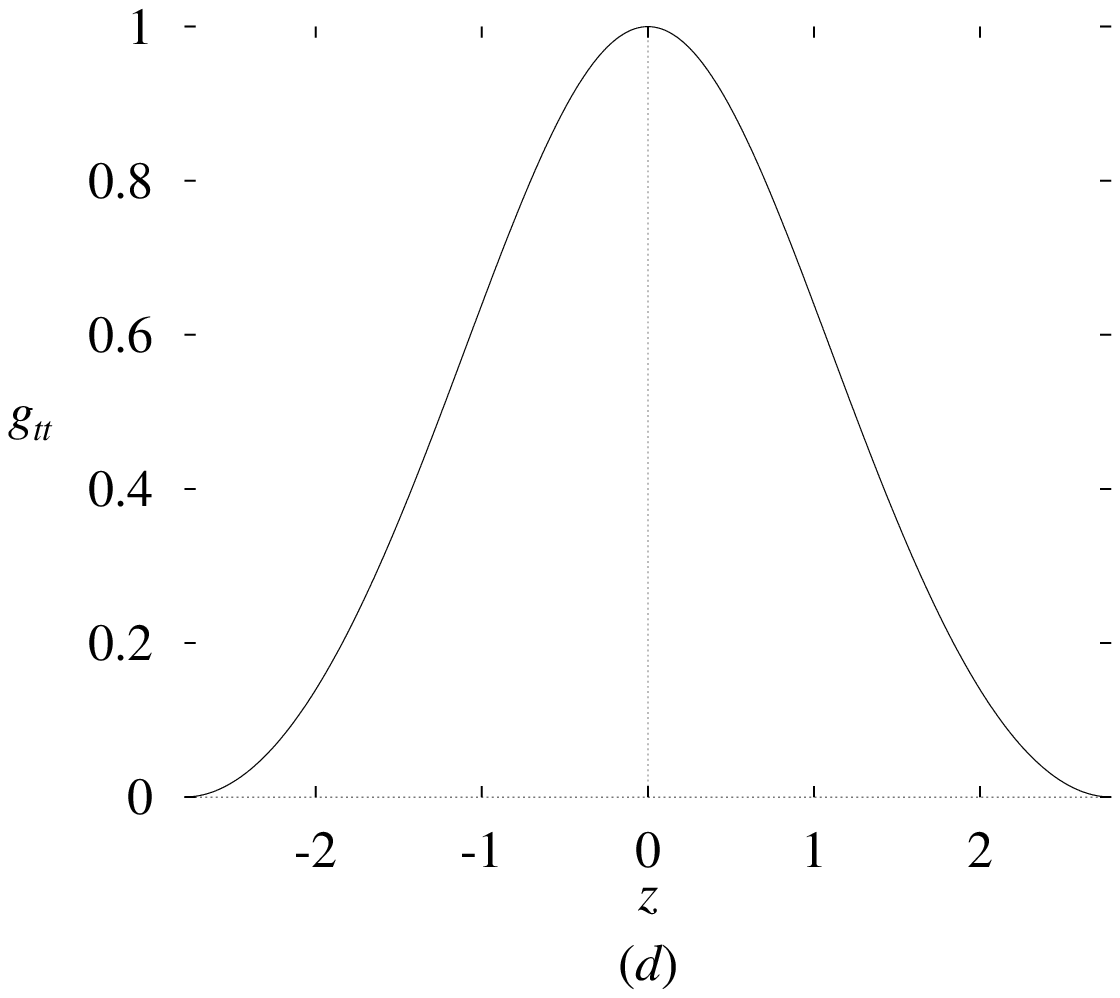,width=7cm}
  \end{center}
  \caption{Solution of the coupled scalar--Einstein equations for $\epsilon =
           0.9$. ({\it a\/})~The Higgs field; ({\it b\/})~The diagonal
           components of the energy-momentum tensor $T^0{}_0 = T^x{}_x =
           T^y{}_y$ and $T^z{}_z$; ({\it c\/})~The function $A(z)$; ({\rm
           d\/})~The metric function $g_{tt}(z) = A^2(z)$.}
  \label{fig:sol}
\end{figure}

For $\epsilon$ varying at this range we can only solve~(\ref{sys2})
numerically, an example for $\epsilon= 0.9$ is shown in figure~\ref{fig:sol}.

For large enough $\epsilon$ the event horizon distance becomes of the same
order as the wall's thickness, $z_{\rm h} = {\rm O}(w)$. In~(\ref{sys2}a,
\ref{sys2c}), the effect of the matter energy momentum tensor is intensified,
which in turn increases the effect of geometry, hence the variation of $A$.
Then, in~(\ref{sys2b}), for a non-singular solution $X$ to exist, $X'$
has to tend to zero faster than $A$ as we approach the horizon. We have then
two distinct possibilities: firstly, the scalar field ignores the geometry and
fluctuates around the false vacuum with an odd parity; the metric in the
wall's core is then approximately de Sitter. Secondly, there is a phase
transition in the behaviour of $X$; that is for some $\epsilon_{\rm max}$
the wall solution ceases to exist and the only non-singular solution turns
out to be exactly the de Sitter one. To put it in a nutshell, either the
field $X$ rolls significantly from the false vacuum or not at all. Note that
when investigating the case of a non-gravitating domain wall in a de Sitter
background, Basu and Vilenkin~\cite{BV} have precisely observed such a phase
transition.  The issue of whether a domain wall can survive in a
Friedmann-Robertson-Walker (FRW) universe has also been analysed in the
case of Euclidean instantons on a de Sitter background including self-gravity
\cite{BV2,BGV}.

Let us now prove that---given the symmetries that we impose to our
solutions---it is the second possibility that holds. To do this, we first
prove that in some range of the parameter $\epsilon$ the de Sitter solution
is unstable to decay into a wall solution; then, we find an estimate of the
value $\epsilon_{\rm max}$ at which domain wall solutions cease to exist.

\subsection*{Stability of the false vacuum-de Sitter solution}

\noindent The de Sitter solution is given by
\bea
  ds^2 &=& \cos^2(kz) \, dt^2 - e^{2kt} \cos^2(kz) \left( dx^2+dy^2 \right) -
           dz^2  \\
  X    &=& 0  \\
  k^2  &=& \epsilon / 3.
\eea
The false vacuum $X = 0$ is an \emph{unstable} solution to wall formation if
there exists a perturbation of $X$, say $\xi(z,t)$, which is an odd function of
$z$ and which is increasing in an unbounded manner with time. Consider then $X
= \Delta \xi(z,t)$, with $\Delta$ an infinitesimally small parameter. We note
that Einstein equations appear in order ${\rm O}(\Delta^2)$ and can be
neglected; the linearised {\it time-dependent\/} field equation to order
$\Delta$ with respect to $\xi$ gives,
\be
  \xi'' - 3k\tan(kz) \xi' - \sec^2(kz) [ \ddot{\xi} + 2k \dot{\xi} ] + 2\xi=0.
\ee
We have a solution given by 
\be
  \xi = e^{k\nu t} \sin kz (\cos kz) ^\nu,
\ee
where
\be
  \nu = -{5\over2} + {1\over2} \sqrt{ 9 +8/k^2}.
\ee
Then $\xi(z,t)$ is an odd function in $z$ and if $\nu>0$ it is exponentially
increasing in time, so that the de Sitter solution $X=0$ is unstable to wall
formation. Now $k^2 = \epsilon / 3$, and we find that for $\epsilon < 3/2$, de
Sitter solution is unstable to wall formation. So it is more energetically 
favorable
for the field $X$ to roll to a domain wall solution than to remain in the false
vacuum de Sitter solution for $\epsilon<3/2$. This is to be expected since the
false vacuum-de Sitter solution is inherently unstable and we have found for
small $\epsilon$ domain wall solutions. Let us now turn to the wall solution.

\subsection*{Existence of wall solutions}

\noindent
As we have already noted, a domain wall solution requires an odd scalar field
$X$ such that $X$ is non-singular ($X'(z_{\rm h}) = 0$) and non-trivial ($X'(0)
> 0$).

Taking the derivative of the scalar equation, we get
\bea
  X''' &=& -3{A' \over A}X''+X'\left[-3{A'' \over A}+3({A' \over A})^2
           + 6X^2 -2 \right]\nonumber \\
       &=& -3{A' \over A}X''+X' F(z) 
\eea
where
\be
  F(z) = \left[ 3\epsilon X^{\prime2} + 3{k^2\over A^2}
         + 6X^2-2 \right]
\ee
and we have used the Einstein equations to replace $A''/A$ and $A'/A$ in
$F(z)$.

At $z=0$ we have,
\be  
  3k^2 = \epsilon[1-X'(0)^2] 
\ee
from the constraint~(\ref{sys2c}); hence
\be
  X'''(0) = X'(0) [ 3\epsilon - 6 k^2 -2] > X'(0) [ \epsilon -2].
\ee
For $\epsilon > 2$, $X'''(0)>0$, so $X''>0$ and $X'$ is increasing away from
$z=0$. Moreover, $F(z)$ is strictly increasing away from $z=0$ and thus $X'$
can never be zero at the horizon.

\begin{figure}[htbp]
  \begin{center}
    \epsfig{file=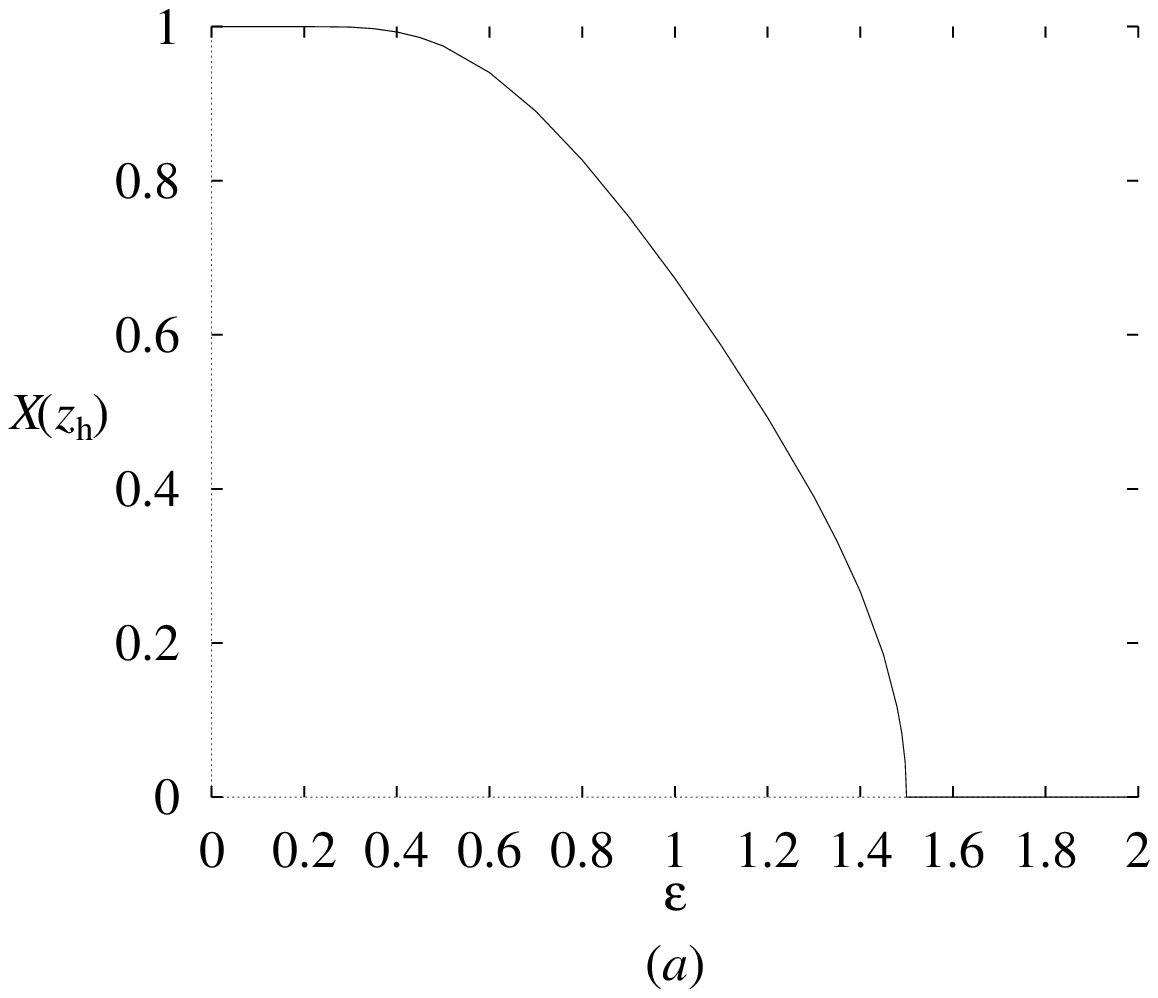,width=7cm} \hfill
    \epsfig{file=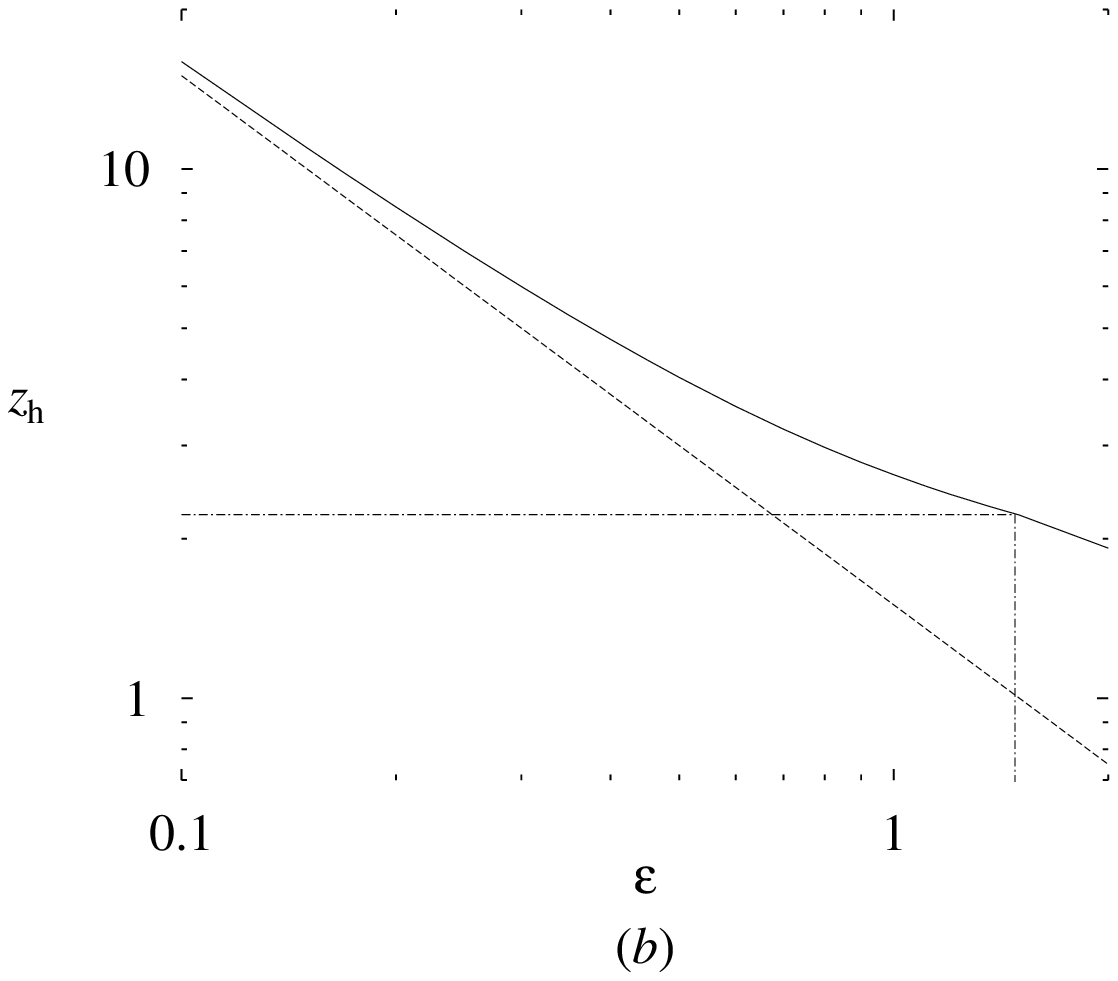,width=7cm}  \\
  \end{center}
  \caption{The phase transition of the Higgs field: ({\it a\/})~shows the value
           of the Higgs field at the wall's horizon, and ({\it b\/})~shows the
           proper distance to the horizon, $z_{\rm h}$, both in function of
           $\epsilon$.}
  \label{fig:pht}
\end{figure}

We can refine numerically this estimate. We find that precisely at $\epsilon =
3/2$ there is a phase transition in the behavior of $X$ (figure~\ref{fig:pht}):
when gravity is very strong the domain wall solution ceases to exist becoming
singular and we are left with the false vacuum-de Sitter solution.

\section*{Concluding remarks}

For a double well potential and $X = X(z)$, we have found:
\begin{itemize}
  \setlength{\itemsep}{0pt}
  \item perturbative solutions for small $\epsilon$,
  \item domain wall solutions for $\epsilon < 3/2$,
  \item that at $\epsilon = 3/2$, $X$ undergoes a second order phase
        transition, and
  \item that for $\epsilon > 3/2$, domain wall solutions do not exist and the
        spacetime is de Sitter. 
\end{itemize}

This phenomenon of phase transition in the behaviour of $X$ is related to the
topology of the gravitating wall (see also \cite{GWG}) and the de Sitter
spaces. Indeed the de Sitter spacetime can be pictured as a four-dimensional
hyperboloid embedded in five-dimensional flat spacetime. Then the spatial
sector of the metric is just the three-sphere which is compact. It turns
out~\cite{BCG} that the domain wall spacetime can also be viewed as a
``squashed'' hyperboloid in flat five-dimensional spacetime. Again,
topologically, this is $S^3$ and the domain wall's space is compactified by
gravity. Therefore, the $t = \mbox{constant}$ slices of spacetime can be
pictured as a squashed ellipsoid with two characteristic lengths: the wall's
width and the distance to the horizon, which varies with $\epsilon$. As gravity
increases, the proper distance to the horizon decreases and becomes eventually
comparable to the wall's width.  Then the phase transition occurs as the two
length scales become identical, and space becomes exactly the de Sitter
three-sphere.

\def\apj#1 #2 #3.{{\it Astrophys.\ J.\ \bf#1} #2 (#3).}
\def\cmp#1 #2 #3.{{\it Commun.\ Math.\ Phys.\ \bf#1} #2 (#3).}
\def\comnpp#1 #2 #3.{{\it Comm.\ Nucl.\ Part.\ Phys.\  \bf#1} #2 (#3).}
\def\cqg#1 #2 #3.{{\it Class.\ Quant.\ Grav.\ \bf#1} #2 (#3).}
\def\jmp#1 #2 #3.{{\it J.\ Math.\ Phys.\ \bf#1} #2 (#3).}
\def\mpla#1 #2 #3.{{\it Mod.\ Phys.\ Lett.\ \rm A\bf#1} #2 (#3).}
\def\ncim#1 #2 #3.{{\it Nuovo Cim.\ \bf#1\/} #2 (#3).}
\def\npb#1 #2 #3.{{\it Nucl.\ Phys.\ \rm B\bf#1} #2 (#3).}
\def\phrep#1 #2 #3.{{\it Phys.\ Rep.\ \bf#1\/} #2 (#3).}
\def\plb#1 #2 #3.{{\it Phys.\ Lett.\ \bf#1\/}B #2 (#3).}
\def\pr#1 #2 #3.{{\it Phys.\ Rev.\ \bf#1} #2 (#3).}
\def\prd#1 #2 #3.{{\it Phys.\ Rev.\ \rm D\bf#1} #2 (#3).}
\def\prl#1 #2 #3.{{\it Phys.\ Rev.\ Lett.\ \bf#1} #2 (#3).}
\def\prs#1 #2 #3.{{\it Proc.\ Roy.\ Soc.\ Lond.\ A.\ \bf#1} #2 (#3).}


\begin{references}
  \bibitem{Vil}     A. Vilenkin, \plb 133 177 1983.
  \bibitem{IS}      J. Ipser and P.Sikivie, \prd 30 712 1984.
  \bibitem{Israel}  W. Israel, \ncim 44B 1 1966.
  \bibitem{RD}      D.Garfinkle and R.Gregory, \prd 41 1889 1990. 
  \bibitem{larryw}  L.M.Widrow, \prd 39 3571 1989. 
  \bibitem{HSF}     C.T. Hill, D.N. Schramm and J.N. Fry, {\it Comment Nucl.
                    Part. Phys.} {\bf 19} 25 (1989).
  \bibitem{Goetz}   G. Goetz, {\it J. Math. Phys.} {\bf 31} 2683 (1990).
  \bibitem{Mukher}  M. Mukherjee, {\it Class. Quantum Grav.} {\bf 10} 131
                    (1993).
  \bibitem{Vil2}    A. Vilenkin, \prl 72 3137 1994 [hep-th/9402085].
  \bibitem{Linde}   A. Linde, \plb 237 208 1994 [astro-ph/9402031].
                    A. Linde and D. Linde, \prd 50 2456 1994 [hep-ph/9402115].
  \bibitem{BCG}     F. Bonjour, C. Charmousis and R. Gregory, [gr-qc/9902081].
  \bibitem{VS}      A. Vilenkin and E.P.S. Shellard, {\it Cosmic strings and
                    other topological defects\/} (Cambridge: Cambridge
                    University Press) 1994.
  \bibitem{BV}      R. Basu and A. Vilenkin, \prd 50 7150 1994 [gr-qc/9402040].
  \bibitem{BV2}     R.Basu and A.Vilenkin, \prd 46 2345 1992.
  \bibitem{BGV}     R.Basu, A.H.Guth and A.Vilenkin, \prd 44 340 1991.
  \bibitem{GWG}     G.W.Gibbons, \npb 394 3 1993.
\end{references}
\end{document}